\documentclass[intlimits,twoside,a4paper]{article}

\usepackage{amsmath,amssymb}
\usepackage{graphicx}

\usepackage[T2A]{fontenc}
\usepackage[cp1251]{inputenc}
\usepackage[eqsecnum]{cmpj2}
\issue{2013}{16}{3}{34001}
\doinumber{10.5488/CMP.16.34001}

\newcommand{\be}{\begin{equation}}
\newcommand{\ee}{\end{equation}}
\newcommand{\bea}{\begin{eqnarray}}
\newcommand{\eea}{\end{eqnarray}}

\articletype{Rapid Communication}

\title[Bath dynamics]%
{Bath dynamics in an exactly solvable qubit model with initial
qubit-environment correlations}
\author[V.V.~Ignatyuk, V.G.~Morozov]{V.V. Ignatyuk\refaddr{label1},
        V.G. Morozov\refaddr{label2}}
\addresses{
\addr{label1} Institute for Condensed Matter Physics
of the National Academy of Sciences of Ukraine,\\
1 Svientsitskii St., 79011 Lviv, Ukraine%
\addr{label2} Moscow State Technical University of Radioengineering, Electronics, and
Automation, \\78 Vernadsky Prospect, 119454 Moscow, Russia}

\date{Received June 16, 2013, in final form August 15, 2013}

\authorcopyright{V.V.~Ignatyuk, V.G.~Morozov, 2013}

\begin{document}

\maketitle

\begin{abstract}
We study the bath dynamics in the dephasing model of a two-state
quantum system (qubit) coupled to an environment of harmonic
oscillators. This model was shown [Morozov et al., Phys. Rev. A, 2012
{\textbf{85}}, 022101] to admit the analytic solution for the
qubit and environment dynamics. Using this solution, we derive the
exact expression for the bath reduced density matrix in the
presence of initial qubit-environment correlations. We obtain the
non-equilibrium phonon distribution function and discuss in detail
the time behavior of the bath energy.
It is shown that only the inclusion of dynamic correlations between the
qubit and the bath ensures the proper time behavior of the quantity which may be
interpreted as the ``environment energy''.

\pacs 03.65.Yz
\end{abstract}

\section{Introduction}

The dynamics of open quantum systems  has attracted a great
deal of interest over the last few decades.
 Its importance is dictated by the prospects of applications in quantum optics,
quantum computing, quantum measurements and control \cite{BP-Book,
Weiss}, as well as by the necessity of a deeper understanding of the
theory itself \cite{Exact-ME,spin-bath,Luczka,Unruh,MMR2012,
MR-CMP2012}. The dynamics of open quantum systems was studied in
several aspects:
  (i) the effect of initial correlations
between an open system and its environment has been investigated in
 \cite{corr1,corr2}; (ii) a new viewpoint concerning the nature and
the measure of non-Markovianity has been presented in
 \cite{non-Markov1, non-Markov2};
 (iii) the effect of
non-equilibrium environment on quantum coherence and the level
populations has been considered in \cite{BD1,BD2}.

The latter problem is of a particular interest. Usually, when
constructing a master equation for the reduced density matrix of
an open system, one considers the bath to be at thermal
equilibrium \cite{BP-Book}, even though there is a build-up of
dynamical correlations \cite{spin-bath,corr1} caused by
the entanglement of quantum states. Moreover, even if the effect of
non-equilibrium environment on the system behavior is not
neglected \cite{BD1}, the intrinsic bath dynamics is beyond
consideration, since the corresponding bath variables are always
integrated out from the equations of motion. Although  this approach
seems to be quite natural as long as one studies solely the open
system dynamics, the investigation of the bath evolution itself can
undoubtedly be an interesting problem,
 yielding some useful hints about how to deal with more realistic
systems, especially with those which possess slow relaxation to
equilibrium and do not admit exact solutions.

In this brief report, we show the relevance of proper (or
intrinsic) bath dynamics. The paper is structured as follows. In
section~2 we derive an exact expression for the bath reduced
density matrix in the so-called dephasing model, describing a
two-state system (qubit) coupled to a bosonic bath
\cite{Luczka,Unruh,MMR2012}. In section~3  the non-equilibrium
distribution function for the bath modes (phonons) is calculated.
 Special attention is paid
to the description of time evolution of the phonon energy
and the correlation energy in the ``qubit-bath'' system. In the
last section we discuss the results and draw final
conclusions.

\section{Bath density matrix in the dephasing model}

We consider a simple version of the spin-boson model describing a
two-state system (qubit) ($S$) coupled to the bath ($B$) of
harmonic oscillators \cite{Luczka,Unruh,MMR2012,MR-CMP2012}. In
the ``spin'' representation for the qubit, the total Hamiltonian
of the model is written as follows (in our units $\hbar = 1$)
\begin{eqnarray}\label{H}
H=H_{\mathrm{S}}+H_{\mathrm{B}}+H_{\mathrm{int}}=\frac{\omega_0}{2}\sigma_3+\sum\limits_k\omega_k
b^{\dagger}_k b_k+\sigma_3\sum\limits_k\left(g_k b^{\dagger}_k+g^*_k
b_k\right),
\end{eqnarray}
where $\omega_0$ is the energy difference between the excited
$|1\rangle$ and the ground $|0\rangle$ states of the qubit, and
$\sigma_3$ is one of the Pauli matrices $\sigma_1$, $\sigma_2$,
$\sigma_3$. Note that the operator $\sigma_3$ can also be written
in the basis of the ground $|0\rangle$ and the excited $|1\rangle$
states of the two-level system as $\sigma_3=|1\rangle\langle
1|-|0\rangle\langle 0|$. Bosonic creation and annihilation
operators $b^{\dagger}_k$ and $b_k$ correspond to the $k$th bath
mode with frequency $\omega_k$, and $g_k$ are the
 coupling constants.

The distinctive feature of the dephasing model (\ref{H}) is
that the average populations of the qubit states do not depend
on time, and hence there is no relaxation to the complete
 equilibrium between the qubit and the environment. In other words,
the model is \textit{non-ergodic\/}\footnote{Another special feature of the model
(\ref{H}) is that the qubit Hamiltonian $H_{\mathrm{S}}$ and the sum $H_{\mathrm{B}}+H_{\mathrm{int}}$ are integrals of motion.}.
Note, however, that the Heisenberg picture operators
 $\sigma_{\pm}(t)=[\sigma_1(t)\pm \ri \sigma_2(t)]/2$  evolve in time,
 leading to a non-trivial decay of  the
 \textit{coherences\/} $\langle\sigma_{\pm}(t)\rangle$. Thus, we have a unique
situation, where the system relaxation may be interpreted physically as
 ``pure'' decoherence and the entropy exchange~\cite{MR-CMP2012}  rather than the energy
dissipation.

It was shown in \cite{MMR2012} that equations of motion for the Heisenberg picture
operators $\sigma_{\pm}(t)$,
$b^{\dagger}_k(t)$, and $b_k(t)$ can be solved exactly with the results
  \begin{eqnarray}
    \label{sig-pm-t}
  \sigma_{\pm}(t)=
   \exp\left\{
    \pm \ri\omega_0 t \mp R(t)
   \right\} \sigma_{\pm}\,,
  \end{eqnarray}
 \begin{eqnarray}
  \label{b(t)}
  b_k(t)=\re^{-\ri\omega_{k}t}\left[b_{k}+\frac{\sigma_{3}}{2}\alpha_{k}(t)
    \right],
&&
 b^{\dagger}_{k}(t)=\re^{\ri\omega_k t}\left[b^{\dagger}_k+\frac{\sigma_{3}}{2}\alpha^*_k(t)
   \right],
 \end{eqnarray}
where
  \begin{eqnarray}
    \label{R_alpha}
R(t)=\sum_{k}\left[\alpha_{k}(t) b^{\dagger}_k-\alpha^*_k(t) b_{k}
\right], && \alpha_k(t)=
2g_{k}\frac{1-\re^{\ri\omega_kt}}{\omega_{k}}\,.
 \end{eqnarray}

We shall use the exact expressions (\ref{sig-pm-t})--(\ref{b(t)})
to evaluate the reduced density matrix of the bath. We  start with
the obvious relation
  \begin{eqnarray}
    \label{rho(t)}
   \rho(t)= \re^{-\ri Ht}\rho(t=0) \re^{\ri Ht}
  \end{eqnarray}
  for the non-equilibrium density matrix of the composite system ($S+B$) and
  assume that the initial density matrix has the form \cite{MMR2012}
  \begin{eqnarray}
   \label{rho-compos-init}
 \rho(t=0)= P_{\psi}\otimes \rho_{\mathrm{B}}(\psi)\equiv|\psi\rangle \langle\psi|\otimes
 \rho_{\mathrm{B}}(\psi),
 \end{eqnarray}
where the projector
 $P_{\psi}=|\psi\rangle \langle\psi|$
  can be expressed
 in terms of the Bloch vector $\vec{v}=\langle\vec{\sigma}\rangle$
 \cite{MMR2012,MR-CMP2012}:
      \begin{eqnarray}
     \label{P-v}
  P^{}_{\psi}=\frac{1}{2}\left(I +
  \vec{v}\cdot\vec{\sigma}\right),
    \end{eqnarray}
where $\langle
\sigma^{}_{3}\rangle=\langle\psi|\sigma^{}_{3}|\psi\rangle$,
$\langle
\sigma^{}_{\pm}\rangle=\langle\psi|\sigma^{}_{\pm}|\psi\rangle$,
and $I$ denotes the $2\times 2$ identity matrix. In
equation~(\ref{rho-compos-init})
$|\psi\rangle=a_0|0\rangle+a_1|1\rangle$  with $|a_0|^2 +
|a_1|^2=1$ is the state vector of the qubit. The
\textit{constrained\/}  initial density matrix of the bath,
$\rho_{\mathrm{B}}(\psi)$, is given by \footnote{ The equation~(\ref{rhoB}) can be
derived by using obvious relations $\sigma_3|1\rangle=|1\rangle$,
$\sigma_3|0\rangle=-|0\rangle$ and the following properties
\cite{MMR2012} of the Hamiltonian (\ref{H}): $\re^{-\beta
H}|0\rangle=\re^{\beta\omega_0/2} \re^{-\beta
H_{\mathrm{B}}^{(-)}}\otimes|0\rangle$, $\re^{-\beta
H}|1\rangle=\re^{-\beta\omega_0/2} \re^{-\beta
H_{\mathrm{B}}^{(+)}}\otimes|1\rangle$.}
   \begin{eqnarray}\label{rhoB}
\rho_{\mathrm{B}}(\psi)\equiv \frac{\langle\psi|\exp(-\beta
H)|\psi\rangle}{\mbox{Tr}_{\mathrm{B}}\langle\psi|\exp(-\beta
H)|\psi\rangle}=\frac{|a_0|^2 \re^{\beta\omega_0/2}\re^{-\beta
H_{\mathrm{B}}^{(-)}}+|a_1|^2 \re^{-\beta\omega_0/2}\re^{-\beta
H_{\mathrm{B}}^{(+)}}}{|a_0|^2 \re^{\beta\omega_0/2}Z_{\mathrm{B}}^{(-)}+|a_1|^2
\re^{-\beta\omega_0/2}Z_{\mathrm{B}}^{(+)}}\,,
  \end{eqnarray}
where $\beta=1/k_{\mathrm{B}} T $. The bath Hamiltonians $H_{\mathrm{B}}^{(\pm)}$ and
the corresponding partition functions $Z_{\mathrm{B}}^{(\pm)}$ are defined as
\begin{eqnarray}\label{HBpm}
H_{\mathrm{B}}^{(\pm)}=\sum\limits_k\omega_k b^{\dagger}_k
b_k\pm\sum\limits_k\left(g_k b^{\dagger}_k+g^*_k b_k\right), &&
Z_{\mathrm{B}}^{(\pm)}=\mbox{Tr}_{\mathrm{B}}\exp\left[-\beta H_{\mathrm{B}}^{(\pm)}\right].
\end{eqnarray}
Physically, the density matrix (\ref{rho-compos-init}) corresponds
to a situation where at times $t < 0$ the open system $S$ is in
thermal equilibrium with its environment $B$, and at time zero one
makes a  perfect (selective) measurement on the system $S$ only.
As a result \cite{BP-Book},
 the system $S$ is prepared in some
pure state $|\psi\rangle $.

Making use of equations~(\ref{rho-compos-init})--(\ref{P-v}) and noting
that $\sigma_3$ is an integral of motion, we can recast
equation~(\ref{rho(t)}) into the form
  \begin{eqnarray}
    \label{rho(t)-1}\nonumber
 & & \hspace*{-45pt} \rho(t)=\frac{1}{2Z^{}_{\mathrm{B}}}
   \left\{
   I+ \langle \sigma^{}_{-}\rangle \sigma^{}_{+}(-t)
    + \langle \sigma^{}_{+}\rangle \sigma^{}_{-}(-t)+\langle \sigma^{}_{3}\rangle\sigma^{}_{3}
   \right\}  \\
 & & \times
      \left\{
  |a_0|^2 \re^{\beta\omega_0/2}\re^{-\beta H_{\mathrm{B}}^{(-)}(-t)}
   +|a_1|^2 \re^{-\beta\omega_0/2}\re^{-\beta H_{\mathrm{B}}^{(+)}(-t)}
  \right\},
  \end{eqnarray}
  where we have introduced the notation
  \begin{equation}
    \label{Z-B}
    Z_{\mathrm{B}}= |a_0|^2 \re^{\beta\omega_0/2}Z_{\mathrm{B}}^{(-)}+|a_1|^2 \re^{-\beta\omega_0/2}Z_{\mathrm{B}}^{(+)}.
  \end{equation}
The bath density matrix is obtained from (\ref{rho(t)-1}) by
taking the trace over the qubit states,
  \begin{eqnarray}
     \label{rhoB(t)}
     \rho_{\mathrm{B}}(t)= \text{Tr}_{\mathrm{S}}\rho(t)=\langle 0|\rho(t)|0\rangle+\langle
     1|\rho(t)|1\rangle.
  \end{eqnarray}
Since $H_{\mathrm{B}}^{(\pm)}(-t)$ does not contain the spin
 operators $\sigma_{\pm} $, the terms with $\sigma_{\pm}(-t)$ do not
contribute to (\ref{rhoB(t)}). Expressing the probabilities
$|a_i|^2$ in terms of $\langle\sigma_3\rangle$ in a usual way, $
|a_1|^2=(1+\langle\sigma_3\rangle)/2$,
$|a_0|^2=(1-\langle\sigma_3\rangle)/2$, it is straightforward to
 manipulate the time-dependent bath density matrix (\ref{rhoB(t)}) to
\begin{eqnarray}
 \label{rho-bath-finalS}\nonumber
\rho_{\mathrm{B}}(t)&=&\frac{1}{4
Z_{\mathrm{B}}}\left\{\left(1-\langle\sigma_3\rangle\right)\re^{\beta\omega_0/2}\left[
\left(1-\langle\sigma_3\rangle\right) \re^{-\beta
H_{\mathrm{B}}^{(-)}}+\left(1+\langle\sigma_3\rangle\right)
\re^{-\beta\left\{H_{\mathrm{B}}^{(+)}-2 \left[H_I(t)-
\Delta\varepsilon_{\mathrm{ph}}(t)\right]\right\}}\right]\right.
\\&&{}+\left.
\left(1+\langle\sigma_3\rangle\right)\re^{-\beta\omega_0/2}\left[
\left(1-\langle\sigma_3\rangle\right) \re^{-\beta\left\{H_{\mathrm{B}}^{(-)}+2
\left[H_I(t)+\Delta\varepsilon_{\mathrm{ph}}(t)\right]\right\}}+\left(1+\langle\sigma_3\rangle\right)
\re^{-\beta H_{\mathrm{B}}^{(+)}}\right] \right\},
\end{eqnarray}
where
\begin{eqnarray}\label{HI(t)}
H_I(t)=\sum\limits_k\left\{g_k \re^{-i\omega_k t}
b^{\dagger}_k+g_k^* \re^{i\omega_k t} b_k \right\}
\end{eqnarray}
denotes a phonon part of $H_{\mathrm{int}}$ in the interaction picture
(with the replacement $t\rightarrow -t$, see
equation~(\ref{rho(t)})), whereas the quantity
\begin{eqnarray}\label{E0}
\Delta\varepsilon_{\mathrm{ph}}(t)=2\sum\limits_k\frac{|g_k|^2}{\omega_k}\left(1-\cos\omega_k
t\right)
\end{eqnarray}
is the non-equilibrium correction to the phonons
energy (see also the next section for discussion).

The expression for the bath density matrix becomes much simpler if
one neglects the initial correlations in the system by taking
a direct product
  \begin{eqnarray}
    \label{rho-comp-uncor}
  \rho(t=0)= P^{}_{\psi}\otimes \rho_{\mathrm{B}}^{(0)}, \qquad \rho_{\mathrm{B}}^{(0)}=\re^{-\beta
  H_{\mathrm{B}}}\big/Z_{\mathrm{B}}^{(0)}, \qquad Z_{\mathrm{B}}^{(0)}=\mbox{Tr}_{\mathrm{B}} \re^{-\beta
  H_{\mathrm{B}}},
  \end{eqnarray}
instead of (\ref{rho-compos-init}). Proceeding in a similar way,
after some algebra one obtains
\begin{equation}
 \label{rho-bath-uncor-final}
\rho_{\mathrm{B}}(t)=\frac{1}{2Z_{\mathrm{B}}^{(0)}}\left\{
\left(1-\langle\sigma_3\rangle\right)\re^{-\beta\left[H_{\mathrm{B}}+\Delta
H_I(t)+\Delta\varepsilon_{\mathrm{ph}}(t)\right]}
+\left(1+\langle\sigma_3\rangle\right)\re^{-\beta\left[H_{\mathrm{B}}-\Delta
H_I(t)+\Delta\varepsilon_{\mathrm{ph}}(t)\right]} \right\},
\end{equation}
where $\Delta H_I(t)=H_I(t)-H_I(0)$ denotes the non-equilibrium
contribution to the correlation energy in the interaction picture.
It is seen from equation~(\ref{rho-bath-uncor-final}) that even
 in this case
 there is a dynamical build-up of correlations in
the system. We touch upon this point in the next
section, when analyzing the relevance of non-equilibrium
correlations.

\section{Phonon non-equilibrium distribution function and energy}

To gain some insight into the time behavior of the bath modes (phonons),
let us first calculate the phonon distribution function $n^{}_{k}(t)$.
This can be done  in two equivalent ways: either using
the exact expression (\ref{rho-bath-finalS}) for the bath density matrix, or
averaging the bosonic Heisenberg picture operators (\ref{b(t)})
over the initial state of the composite system. Here, we shall follow the latter
 procedure which is simpler. Taking the initial density matrix
 in the form  (\ref{rho-compos-init}) and then applying
 the unitary transformation technique \cite{MMR2012}
 (or the method of the displaced harmonic oscillator modes
\cite{PRA2013}), after some straightforward algebra one obtains
\begin{eqnarray}\label{nk(t)}
  n_k(t)\equiv
  \mbox{Tr}_{\mathrm{S,B}}\left\{\rho(t=0)b^{\dagger}_k(t)b_k(t)\right\}=n_k(0)+\frac{2|g_k|^2}{\omega_k^2}
(A(\psi)\langle\sigma_3\rangle +1)(1-\cos\omega_k t),
\end{eqnarray}
 where
 \begin{eqnarray}\label{nkt0} n_k(t=0)=
 [\exp(\beta\omega_k)-1]^{-1}+|g_k|^2/\omega_k^2
\end{eqnarray}
is the initial phonon distribution function, and the function
 \begin{eqnarray}\label{A}
A(\psi)=\frac{\sinh(\beta\omega_0/2)-\langle\sigma_3\rangle\cosh(\beta\omega_0/2)}
{\cosh(\beta\omega_0/2)-\langle\sigma_3\rangle\sinh(\beta\omega_0/2)}
\end{eqnarray}
represents the contribution of initial correlations.

With equations~(\ref{nk(t)}) and (\ref{nkt0}), it is easy to calculate
the time evolution of the non-equilibrium phonon energy
$\varepsilon(t)=\sum_k \omega_k n_k(t)$. The final result is
conveniently written in terms of
  the bath spectral density $J(\omega)$ which is introduced by the
well-known rule \cite{BP-Book,Weiss,MMR2012}
 \begin{equation}
 \label{J}
 \sum\limits_k 4|g_k|^2
 f(\omega_k)=\int\limits_0^{\infty} J(\omega) f(\omega)\rd\omega.
 \end{equation}
After simple manipulations we arrive at
 \begin{equation}
    \label{eps}
    \varepsilon(t)=\varepsilon(t=0)+\frac{1}{2}\left(A(\psi)\langle\sigma_3\rangle+1\right)
    \int\limits_0^{\infty}\frac{J(\omega)}{\omega}
   \left(1-\cos\omega t\right)\rd\omega,
 \end{equation}
where the initial phonon energy is given by
 \begin{equation}
 \label{epst0}
\varepsilon(t=0)=\varepsilon_0+\Delta\varepsilon=
\sum\limits_k\omega_k[\exp(\beta\omega_k)-1]^{-1}+\frac{1}{4}\int\limits_0^{\infty}
\frac{J(\omega)}{\omega}\,\rd\omega.
 \end{equation}
 Here the last term occurs due to initial correlations in the
  system.

 Usually
\cite{BP-Book,Weiss,MMR2012}, the spectral density function is
chosen in the form
\begin{eqnarray}\label{J1}
J(\omega)=\lambda_s\Omega^{1-s}\omega^s\exp(-\omega/\Omega),
\end{eqnarray}
where $s>0$ and $\lambda_{s}$ is a dimensionless coupling constant.
This formula ensures both a proper low-frequency behavior of $J(\omega)$
 and a cut-off at high frequencies ($\omega\gg\Omega$).
 The case $s = 1$ is usually called the ``Ohmic'' case, the case
$s > 1$ ``super-Ohmic'', and the case $0 < s < 1$ ``sub-Ohmic''.
Using expression (\ref{J1}), it is possible to analyze the time behavior of
 $\varepsilon(t)$ for different $s$,  but in this brief report
 we would like to discuss only one physically interesting point related to the
  result (\ref{eps}) for the phonon energy.

At first glance, the fact that the phonon energy (\ref{eps})
depends on time may appear as an apparent paradox. Indeed, on the one hand,
 one may conclude that there is an energy exchange between the qubit and the bath.
 On the other hand,  the qubit Hamiltonian $H_{\mathrm{S}}$ commutes with the total
  Hamiltonian (\ref{H}), and hence the qubit energy $\langle H_{\mathrm{S}}\rangle$ does not depend
   on time. To explain this paradox, let us calculate the non-equilibrium
   \textit{correlation energy\/} $\varepsilon_{\mathrm{cor}}(t)=\langle H^{}_{\mathrm{int}}(t)\rangle$,
    where  $H^{}_{\mathrm{int}}(t)$ is the interaction term in equation~(\ref{H})
    (in the Heisenberg picture) and the average is taken over the initial state
     (\ref{rho-compos-init}). After some algebra, which we omit, we obtain
   \begin{equation}
   \label{epsCor}
  \varepsilon_{\mathrm{cor}}(t)\equiv\mbox{Tr}_{\mathrm{S,B}}\left\{\rho(t=0)
  H_{\mathrm{int}}(t)\right\}= \varepsilon_0+ \left(2
   A(\psi)\langle\sigma_3\rangle+1\right)\Delta\varepsilon-\varepsilon(t).
  \end{equation}
 Combining this expression with equation~(\ref{eps}), it is easy to check that
  the sum  $\varepsilon(t)+ \varepsilon_{\mathrm{cor}}(t)$ is a time-independent
quantity (see also a footnote on page 2). We see that
 the dynamics of the correlation energy $\varepsilon_{\mathrm{cor}}(t)$
exactly compensates the time dependence of the non-equilibrium phonon energy
$\varepsilon(t)$, ensuring the energy conservation law.
 Physically, the sum  $\varepsilon(t)+ \varepsilon_{\mathrm{cor}}(t)$ is precisely
 the quantity which should be interpreted as the \text{environment energy\/}.

One more remark is to the point. It can be seen from
 equations~(\ref{A}), (\ref{eps}), and (\ref{epsCor}) that both the non-equilibrium phonon
energy $\varepsilon(t)$ and the correlation energy
$\varepsilon_{\mathrm{cor}}(t)$ do not depend on time under conditions
$\langle\sigma_3\rangle=\pm 1$. Note in this connection that the
correlational contribution $\gamma_{\mathrm{cor}}(t)$ to the generalized
decoherence function \cite{MMR2012} vanishes for the same values of
the mean inversion population of the levels, manifesting a close
relationship between the essentially non-equilibrium behavior of
the correlation energy and the onset of the additional channel of
decoherence in the system.

\section{Conclusions}

Here, we present a summary of the results and discuss their relation to some
problems in the dynamics of open quantum systems.

We have derived exact expressions (\ref{rho-bath-finalS})
 and (\ref{rho-bath-uncor-final}) for the bath density matrix in the
 model (\ref{H}) which describes the dephasing mechanism of decoherence
 in a qubit interacting with a bosonic environment.
To the best of our knowledge,
 a derivation of a bath density matrix has never been performed in the theory
of open quantum systems. The explicit form of $\rho_{\mathrm{B}}(t)$
could be essential, for instance, when constructing master equations (especially
non-Markovian) and taking into account the intrinsic dynamics
of the environment along with the equation of motion for
$\rho_{\mathrm{S}}(t)$. Such an approach would modify the well-known
Zwanzig-Nakajima projection technique
\cite{BP-Book,Exact-ME,corr1} where the bath degrees of
 freedom are ``eliminated''. We believe that this modification  is quite natural
 in the case of the finite size of the
bath, when all the environmental modes are involved in the
composite system dynamics, and a back-flow of  energy
(information) from the bath to the open system is essential. Thus, the
exact solutions (\ref{rho-bath-finalS}) and
(\ref{rho-bath-uncor-final}) can give a valuable insight into general
properties of the dynamics of decoherence and can serve as a
step toward consistent
derivation of master equations ensuring regular behavior of
composite systems on all timescales and for strong coupling
regimes.

 Our analysis of the phonon energy in section~3 illustrates
  the special role of dynamic correlations between an open system and its
  environment. We have seen that a ``naive'' picture  with the
  ``energy exchange between the qubit and the bath'' is inadequate
   (even in the case of weak coupling),
  and only the proper inclusion of non-equilibrium correlations ensures the
  conservation of the total energy.

%
%

\ukrainianpart
\title{Динаміка термостату для точної моделі кубіту при наявності початкових кореляцій з його оточенням}
\author{В.В.~Ігнатюк\refaddr{label1}, В.Г.~Морозов\refaddr{label2}}
\addresses{
\addr{label1} Інститут фізики конденсованих систем НАН України,
вул. Свєнціцького, 1, 79011 Львів, Україна
\addr{label2} Московський державний технічний університет
радіоелектроніки та автоматики, \\ просп. Вернадського, 78, 119454 Москва, Росія}
%
%
%

\makeukrtitle

\begin{abstract}
\tolerance=3000%
Проведено дослідження динаміки термостату у випадку моделі з
розфазуванням, що описує дворівневу квантову систему (кубіт), яка
взаємодіє з гармонічними осциляторами зі свого оточення. Ця модель
має аналітичний розв'язок [Morozov et al., Phys. Rev. A, 2012
{\textbf{85}}, 022101] як для спінових змінних, так і для змінних
термостату. Використовуючи цей розв'язок, отримано аналітичний вираз для приведеної матриці густини термостату
при наявності початкових кореляцій. Отримано нерівноважну функцію розподілу фононів та
детально досліджено часову еволюцію енергії термостату.
Показано, що лише належне врахування динамічних кореляцій між кубітом та його оточенням забезпечує збереження
величини, яку слід вважати ``енергією оточення''.

\keywords квантові відкриті системи, кореляції, кубіт

\end{abstract}

\end{document}